\documentstyle[12pt]{article}
\title{\bf Acceleration of the universe in matter dominant era by conformal symmetry breaking}
\author{
  F. Darabi\thanks{e-mail: f-darabi@azaruniv.edu}
\\
{\small Department of Physics, Azarbaijan Shahid Madani University, Tabriz,
53714-161,  Iran.}}
\begin{document}
\maketitle
\begin{abstract}
We introduce a scenario in which the breakdown of conformal symmetry is responsible for the acceleration of universe in the matter dominant era. In this regard, we consider a self interacting scalar field non-minimally coupled to the
Ricci scalar and the trace of energy-momentum tensor. For a traceless energy-momentum tensor in radiation dominant era, the coupling to matter vanishes and we are left with a conformal invariant gravitational action of Deser, where the universe may experience a decelerating phase in agreement with observations. In matter dominant era, the coupling to matter no longer vanishes, the conformal symmetry is broken down, and the matter inevitably becomes pressureless. The corresponding field equations are obtained and it is shown that the universe may have an accelerating phase in this era, provided that the value of self interaction coupling constant satisfies an specific lower bound. Moreover, we provide a reasonable solution to the coincidence problem.
\end{abstract}
\vspace{2cm}
\section{Introduction}

The recent cosmological observations obtained by SNe Ia {\cite{c1}}, WMAP {\cite{c2}}, SDSS {\cite{c3}} and X-ray {\cite{c4}} show that the universe is currently experiencing an unexpected accelerated expansion. These observations also indicate that our universe, consisting of about $70 \%$ dark energy, $30\%$ dust matter, and negligible radiation, is almost spatially flat. To account for the cosmic acceleration, many theories and models have been proposed
based on the assumption that the dark energy is the main cause for the acceleration
of universe. One natural candidate for the dark energy is a tiny positive cosmological constant. However, this suffers from the well known cosmological constant and the coincidence problems \cite{4}.
Alternative proposals to explain the acceleration of universe, without the coincidence problem, are based on the dynamical dark energy models such as quintessence \cite{quint}, phantom \cite{phant}, quintom \cite{quintom}, and interacting models of dark energy \cite{Jamil}. Also, the holographic principle of quantum gravity theory has recently introduced a framework which may simultaneously provide a solution to both dark energy and the coincidence problems \cite{holoprin}.

Extended theories of gravity \cite{mauro}-\cite{defelice}, on the other hand, try to address the problem of acceleration of the universe
without resorting to dark energy. The idea of extended theories of gravity is adding physically motivated higher order curvature invariants and non-minimally coupled scalar fields to the action \cite{odi,farh}. Modified theories
of gravity, take into account an effective action where the gravitational Lagrangian is generic functions $f(R)$ \cite{Review-N-O}-\cite{Clifton:2011jh}, $f(G)$ \cite{15}-\cite{15-5}, $f(T)$ \cite{f(T)-Refs}-\cite{Local-Lorentz-invariance}, or their combinations \cite{F(GR)-gravity}-\cite{F(GR)-gravity2}. 

The breakdown of conformal symmetry in cosmology for scalar fields has been used many times, see for example \cite{Grib}, \cite{Bekenstein}. A general feature of conformally invariant theories is the presence of varying dimensional coupling constants. Hence, the introduction of a constant dimensional parameter into these theories breaks down the conformal symmetry and a preferred conformal frame, in which the dimensional parameter has the assumed constant configuration, is singled out. Conformal symmetry may be broken down by defining a preferred conformal frame in terms of the large scale properties of a finite universe.
The breakdown of conformal symmetry then becomes a framework in which one may look for the origin of the gravitational coupling of matter and the
cosmological constant. Deser, was the first who achieved this goal \cite{Deser}. In a previous paper, we have shown that the idea of conformal symmetry breaking has the capability of accounting for the acceleration of universe in the broken phase of a conformal invariant gravitational model \cite{Darabi}. The purpose of present paper is to assert on this capability by showing that a generalized version of Deser's model, consisting of a self-interacting scalar field coupled to matter as well as gravity, may be used to explain the accelerating behavior of universe in the pressureless matter dominant era and provide a reasonable solution to the coincidence problem.

\section{Conformal symmetry and its breakdown in gravitational models}

\subsection{Scalar field non-minimally coupled to gravity}

In this section, we review briefly the work of Deser \cite{Deser}. Let us
consider the action\footnote{We use the metric sign convention $g_{\mu \nu}=diag(+,-,-,-)$.} 
\begin{equation}
S[\phi]=\frac{1}{2} \int \!d^4 x \sqrt{-g} (g^{\mu \nu} \nabla_{\mu} \phi
\nabla_{\nu} \phi +\frac{1}{6} R \phi^2), 
\label{1}
\end{equation}
which describes a non-minimally coupled real scalar field $\phi$ to gravity
described by the scalar curvature $R$. Varying the action with respect
to $\phi$ and $g_{\mu \nu}$ results in the following equations
\begin{equation}
(\Box -\frac{1}{6} R)\phi=0,
\label{2}
\end{equation}
\begin{equation}
G_{\mu \nu}=6\phi^{-2} \tau_{\mu \nu}(\phi),
\label{3}
\end{equation}
where $G_{\mu \nu}$ is the Einstein tensor and the right hand side reads
as
\begin{equation}
\tau_{\mu \nu}(\phi)= - [\nabla_\mu \phi \nabla_\nu \phi - \frac{1}{2}g_{\mu \nu}
\nabla_\alpha \phi \nabla^\alpha \phi] -
\frac{1}{6}(g_{\mu \nu}\Box -\nabla_\mu \nabla_\nu)\phi^2. 
\end{equation}
Taking the trace of (\ref{3}) results in
\begin{equation}
\phi(\Box -\frac{1}{6} R)\phi=0, 
\label{5}
\end{equation}
which is consistent with equation (\ref{2}). This consistency is a consequence of the conformal invariance of action (\ref{1}) under the following conformal transformations
\begin{equation}
\phi \rightarrow \bar{\phi}=\Omega^{-1}(x) \phi ,\:\:\:\:\:\:\:\:\:\:
g_{\mu \nu}\rightarrow \bar{g}_{\mu \nu}=\Omega^2 (x) g_{\mu \nu}, 
\label{6}
\end{equation}
where $\Omega(x)$ is the conformal factor, being an arbitrary and smooth
function of space-time. Now, we add a matter source $S_{m}$ to (\ref{1}) as
\begin{equation}
S = S[\phi] + S_{m}, 
\label{7}
\end{equation}
so that the field equations become
\begin{equation}
(\Box -\frac{1}{6} R)\phi=0, 
\label{8}
\end{equation}
\begin{equation}
G_{\mu \nu}=6\phi^{-2}[\tau_{\mu \nu}(\phi)+T_{\mu \nu}], 
\label{9}
\end{equation}
where $T_{\mu \nu}$ is a $\phi$-independent matter energy-momentum tensor. By comparing the trace of (\ref{9}) with Eq.(\ref{8}), an algebraic requirement emerges as 
\begin{equation}
T_\mu ^{\mu}=T=0. 
\label{10}
\end{equation}
This implies that in order for a matter source can couple consistently to such a conformal invariant gravitational models, it should be traceless. To breakdown the conformal symmetry one may add a dimensional mass 
term $\frac{1}{2}\int\!d^4 x \sqrt{-g} \mu^2 \phi^2$ to the action
(\ref{7}), with $\mu$ being a constant mass parameter. In fact, the conformal invariance breaks down when a particular conformal frame is chosen in which the dimensional parameter $\mu$ takes on a constant configuration. The choice of conformal frame is usually suggested by the physical considerations. The presence of above mass term in the action leads to
\begin{equation}
(\Box -\frac{1}{6} R-\mu^2)\phi=0, 
\label{13}
\end{equation}
\begin{equation}
G_{\mu \nu}-3\mu^2 g_{\mu \nu}=6\phi^{-2} [\tau_{\mu \nu}(\phi)+T_{\mu \nu}],
\label{14}
\end{equation}
\begin{equation}
\mu^2 \phi^2=T.
\label{12}
\end{equation}
One may determine a conformal frame considering the large scale properties of the observed universe. In this way, one may take $\mu^{-1}$ as the characteristic
size of the universe $a_0$ and $T$ as the average density of the large scale distribution of matter $ M a_0^{-3}$, $M$ being the observed mass of the universe. As a consequence of (\ref{12}), this leads to the estimation of the constant background value of $\phi$ as  
\begin{equation}
{\phi}^{-2} \sim a_0^{-2}(M/a_0^3)^{-1} \sim a_0/M \sim G,
\label{15}
\end{equation}
where the empirical cosmological relation $GM/a_0 \sim 1$
has been used \cite{Deser}. Substituting the background value of $\phi\sim
G^{-2}$ into the field equations (\ref{13}) and (\ref{14}) gives respectively,
an identity $0=0$ and the standard Einstein equation
\begin{equation}
G_{\mu \nu}= 8\pi G T_{\mu \nu}+3\mu^2 g_{\mu \nu},
\label{16}
\end{equation}
where the term $3\mu^2$ appears as an effective positive cosmological constant $\Lambda$ with the correct order of magnitude $\sim a_0^{-2}$.

\subsection{Self-interacting scalar field non-minimally coupled to gravity }

Now, we aim to generalize the Deser's model to include a conformal invariant
self-interacting scalar potential and redo the previous investigation. 
In this regard, we consider the scalar field action
\begin{equation}\label{Deser}
\bar{S}[\phi]=\frac{1}{2} \int \!d^4 x \sqrt{-g} (g^{\mu \nu} \nabla_{\mu} \phi
\nabla_{\nu} \phi +\frac{1}{6} R \phi^2-\frac{1}{4}\lambda \phi^4), 
\end{equation}
where $\lambda$ is a dimensionless self-interacting coupling constant. The modified field equations corresponding to the action $S = \bar{S}[\phi] + S_{m}$ are immediately obtained, respectively as 
\begin{equation}
(\Box -\frac{1}{6}R+\frac{1}{2}\lambda \phi^2)\phi=0,
\label{2'}
\end{equation}
\begin{equation}
G_{\mu \nu}=6\phi^{-2}[\bar{\tau}_{\mu \nu}(\phi)+T_{\mu \nu}],
\label{3'}
\end{equation}
where 
\begin{equation}
\bar{\tau}_{\mu \nu}(\phi)= {\tau}_{\mu \nu}(\phi)-\frac{1}{8}g_{\mu \nu}\lambda \phi^4. 
\end{equation}
Equations (\ref{2'}) and (\ref{3'}) become consistent provided $T=0$. 
As in the Deser's model, the corresponding conformal symmetry is broken down by adding the mass term $\frac{1}{2}\int\!d^4 x \sqrt{-g} \mu^2 \phi^2$ which leads to $\mu^2 \phi^2=T$. Then, the equations (\ref{2'}) and (\ref{3'}) become conformally non-invariant and in the specific cosmological conformal frame defined by (\ref{15}), we obtain the Einstein equation
\begin{equation}
G_{\mu \nu}= 8\pi G T_{\mu \nu}+\bar{\Lambda} g_{\mu \nu},
\end{equation}
where $\bar{\Lambda}=3\bar{\mu}^2=3(\mu^2+\frac{3 \lambda}{16\pi G})$. 
Therefore, the conformaly invariant self interacting potential may contribute a (positive/negative) constant to the cosmological constant corresponding
to (positive/negative) value of $\lambda$.

\section{Dynamics of universe in radiation dominant era}

\subsection{Scalar field non-minimally coupled to gravity}

Now, we study the evolution of universe before symmetry breaking, namely at the radiation dominant era. Let us consider the action in the radiation dominant era $S = S[\phi] + S_{r}$ where $S_{r}$ is the action corresponding to the radiation. We take $g_{\mu \nu}$ as the flat ($k=0$) Friedmann-Robertson-Walker metric
\begin{eqnarray}\label{metric}
ds^2=dt^2-a^2(t)[dr^2+r^2(d\theta^2+\sin^2\theta
d\phi^2 )],
\end{eqnarray}
and assume $T^{~r}_{\mu \nu}$ as the perfect fluid describing the radiation
by $T_\mu ^{\mu~r}=0$. The field equations (\ref{8}), (\ref{9}) yield
\begin{equation}
\frac{\dot{a}^2}{a^2}+\frac{\dot{\phi}^2}{\phi^2}+2\frac{\dot{a}}{a}\frac{\dot{\phi}}{\phi}=2\frac{\rho_{r}}{\phi^2},\label{24'}
\end{equation}
\begin{equation}
2\frac{\ddot{a}}{a}+\frac{\dot{a}^2}{a^2}=-6\frac{p_{r}}{\phi^2},\label{25'}
\end{equation}
\begin{equation}\label{26}
\frac{\ddot{\phi}}{\phi}+\frac{\ddot{a}}{a}+3\frac{\dot{a}}{a}\frac{\dot{\phi}}{\phi}
+\frac{\dot{a}^2}{a^2}=0,
\end{equation}
where $a$ is the scale factor, $\rho_{r}$ and $p_{r}$ are the density and pressure of the radiation respectively, and $\dot{}$ means derivative with respect to the cosmological time $t$.
The acceleration equation is obtained by combining Eqs.(\ref{24'}),  (\ref{25'}) as follows 
\begin{equation}
\frac{\ddot{a}}{a}=\frac{\dot{\phi}^2}{2{\phi}^2}+\frac{\dot{a}}{a}\frac{\dot{\phi}}{\phi}
-\frac{1}{{\phi}^2}(\rho_{r}+3p_{r}).\label{26'}
\end{equation}
Putting the power law behaviors $\rho_{r}=A a^{\alpha}$, $\phi=B a^{\beta}$
and $a=Ct^{\gamma}$ with the equation of state $p_{r}=\frac{1}{3} \rho_{r}$ (for radiation) into Eq.(\ref{26'}) we obtain 
\begin{equation}
\gamma=\frac{2}{2\beta-\alpha}.\label{28}
\end{equation}
It is easily seen that the universe may experience a decelerating phase in
radiation dominant era, in agreement with observations, provided $\gamma<1$ which means
\begin{equation}
\beta>1+\frac{\alpha}{2}.\label{31}
\end{equation}

\subsection{Self-interacting scalar field non-minimally coupled to gravity }

The field equations corresponding to the action $S = \bar{S}[\phi] + S_{r}$
are obtained
\begin{equation}
\frac{\dot{a}^2}{a^2}+\frac{\dot{\phi}^2}{\phi^2}+2\frac{\dot{a}}{a}\frac{\dot{\phi}}{\phi}-\frac{1}{4}\lambda
\phi^2=2\frac{\rho_{r}}{\phi^2},\label{24''}
\end{equation}
\begin{equation}
2\frac{\ddot{a}}{a}+\frac{\dot{a}^2}{a^2}-\frac{3}{4}\lambda
\phi^2=-6\frac{p_{r}}{\phi^2},\label{25''}
\end{equation}
\begin{equation}\label{26''}
\frac{\ddot{\phi}}{\phi}+\frac{\ddot{a}}{a}+3\frac{\dot{a}}{a}\frac{\dot{\phi}}{\phi}
+\frac{\dot{a}^2}{a^2}-\frac{1}{2}\lambda \phi^3=0.
\end{equation}
The acceleration equation is obtained as 
\begin{equation}
\frac{\ddot{a}}{a}=\frac{\dot{\phi}^2}{2{\phi}^2}+\frac{\dot{a}}{a}\frac{\dot{\phi}}{\phi}
+\frac{1}{4}\lambda \phi^2-\frac{1}{{\phi}^2}(\rho_{r}+3p_{r}).\label{26'''}
\end{equation}
Using the power law solutions $\rho_{r}=A a^{\alpha}$, $\phi=B a^{\beta}$
and $a=Ct^{\gamma}$ together with the equation of state $p_{r}=\frac{1}{3} \rho_{r}$ into Eq.(\ref{26'''}) we obtain 
\begin{equation}
\gamma=\frac{2}{2\beta-\alpha}\:\:\:,\:\:\: \gamma \beta=-1\:\:\:\Rightarrow \:\:\:\alpha=4\beta.\label{28''}
\end{equation}
Therefore, the universe may experience a decelerating phase in
radiation dominant era provided $\gamma<1$ which means
\begin{equation}
\alpha<-4\:\:\:,\:\:\:\beta<-1.\label{31'}
\end{equation}

\section{Dynamics of universe in matter dominant era}

\subsection{Scalar field non-minimally coupled to gravity}

The decelerating phase of universe continues until a cosmological phase transition occurs from radiation dominant era to matter dominant era having a dominant source of visible matter with vanishing pressure. Then, the action is $S = S[\phi] + S_{m}$ and the energy-momentum tensor corresponding
to $S_{m}$ reads as 
\begin{equation}
{T}^{~m}_{\mu \nu}=(\rho_{m}){u}_{\mu} {u}_{\nu}, \label{23'}
\end{equation}
with a non-vanishing trace 
\begin{equation}\label{23''}
T^{m}=\rho_{m}.
\end{equation}
According to the discussion in subsection 2.1, the field equations (\ref{8}), (\ref{9}) are not consistent unless a mass term is added to the action as
\begin{equation}
S = S[\phi] + \frac{1}{2}\int\!d^4 x \sqrt{-g} \mu^2 \phi^2+S_{m}, 
\label{7'''}
\end{equation}
which obviously breaks down the conformal symmetry. It is worth mentioning
that, the mass term in Deser's {\it gravitational} model is added by hand to provide a consistent coupling of matter (with non-vanishing trace) to gravity, and also account for the gravitational and cosmological constants. However, in the present {\it cosmological} framework, in order for the consistency equation $\mu^2 \phi^2=T^m$ holds, the mass term is naturally
introduced due to the cosmological dominance of matter over radiation in the matter dominant era which introduces a non-vanishing trace. If the cosmological conformal symmetry breaking occurs naturally during the phase transition from radiation dominance to matter dominance, then the important consequence of this symmetry breaking is the appearance of gravitational coupling $G\sim\phi^{-2}$, and positive cosmological constant $\Lambda\sim\mu^2$ in the Einstein equation \begin{equation}
G_{\mu \nu}= 8\pi G T_{\mu \nu}+ \Lambda g_{\mu \nu}.
\label{16'}
\end{equation}
The induced positive cosmological constant $\Lambda$ resulting
from cosmological conformal symmetry breaking in the matter dominant era becomes a potential candidate for dark energy responsible for the acceleration of universe.

\subsection{Self-interacting scalar field non-minimally coupled to gravity }

Like the case of scalar field non-minimally coupled to gravity, in the 
case of self-interacting scalar field, the decelerating phase of universe continues until a phase transition happens from radiation dominant to matter dominant eras. The corresponding action is $S = \bar{S}[\phi] + S_{m}$ and the energy-momentum tensor corresponding to $S_{m}$ is the same as (\ref{23'})
with a non-vanishing trace $T^{m}=\rho_{m}$.
The field equations (\ref{2'}) and (\ref{3'}) are not consistent unless a mass term is added to the action as
\begin{equation}
S = \bar{S}[\phi] + \frac{1}{2}\int\!d^4 x \sqrt{-g} \mu^2 \phi^2+S_{m}, 
\label{4}
\end{equation}
which breaks down the conformal symmetry. The resulting Einstein equation
is given by \begin{equation}
G_{\mu \nu}= 8\pi G T_{\mu \nu}+ \bar{\Lambda} g_{\mu \nu}.
\label{11}
\end{equation}
As in the case of scalar field non-minimally coupled to gravity, the induced positive cosmological constant $\bar{\Lambda}$, resulting from cosmological breakdown of conformal symmetry in the matter dominant era, becomes a candidate
for dark energy and responsible for the acceleration of universe.

\subsection{Coincidence problem}

It is so difficult to understand why we happen to live in the special epoch where $\rho_V \sim \rho_m$. This is known as {\it coincidence problem} \cite{Carr}.
We show that the cosmological breakdown of conformal symmetry in both scalar
field and self interacting scalar field is capable of shedding light on the coincidence problem. In the case of scalar field non minimally coupled to
gravity, the energy density of vacuum is given by
\begin{equation}
\rho_V=\frac{\Lambda}{8 \pi G},\label{33}
\end{equation}
whereas according to the model of Deser we have 
\begin{equation}
\rho_V\sim \frac{\mu^2}{G}.\label{34}
\end{equation}
On the other hand, considering $\phi^2\sim G^{-1}$ and Eqs.(\ref{12}) and (\ref{23''}) we obtain
\begin{equation}\label{35}
\mu^2 \sim G\rho_m.
\end{equation}
Finally, combination of (\ref{34}) and (\ref{35}) leads to the desired coincidence
\begin{equation}\label{36}
\rho_V \sim \rho_m.
\end{equation}
This is also the case for the model of self interacting scalar
field, where $\mu^2$ is replaced by $\bar{\mu}^2$. 

\section{Self interacting scalar field non-minimally coupled to gravity and matter}

\subsection{Conformal symmetry and its breakdown}

Now, we generalize the model of ``self interacting scalar field non-minimally coupled to gravity'' to the model of ``self interacting scalar field non-minimally coupled to gravity and matter''. In so doing, we consider the following action
\begin{equation}\label{37}
\tilde{S}[\phi]=\frac{1}{2} \int \!d^4 x \sqrt{-g} (g^{\mu \nu} \nabla_{\mu} \phi\nabla_{\nu} \phi +\frac{1}{6} R \phi^2-\frac{1}{4}\lambda T \phi^4), \end{equation}
where $T$ is the trace of energy-momentum tensor and $\lambda$ is necessarily a dimensional self-interacting coupling constant. This action is conformally invariant
provided that $T=0$, and for any matter source with $T^m\neq0$, the conformal
symmetry is broken down. This is because, $\lambda$ is a constant dimensional parameter and so the non-vanishing term $\lambda T^m \phi^4$ in this action plays just the same role of {\it constant mass} term in Deser's model, hence this term does not properly transform to keep the conformal symmetry of the whole action. The generalized field equations corresponding to the action $S = \tilde{S}[\phi] + S_{m}$ are obtained, respectively as 
\begin{equation}
(\Box -\frac{1}{6} R+\frac{1}{2}\lambda T^m \phi^2)\phi=0,
\label{38}
\end{equation}
\begin{equation}
G_{\mu \nu}=6\phi^{-2}[\tilde{\tau}_{\mu \nu}(\phi)+T^m_{\mu \nu}],
\label{39}
\end{equation}
where 
\begin{equation}\label{40}
\tilde{\tau}_{\mu \nu}(\phi)= {\tau}_{\mu \nu}(\phi)-\left[\frac{1}{2}T^m g_{\mu \nu}-(\rho_m+p_m)u_{\mu}u_{\nu}\right]\lambda \phi^4. 
\end{equation}
The trace of Eq.(\ref{39}) is obtained as
\begin{equation}
(\Box -\frac{1}{6} R+\frac{1}{2}\lambda T^m \phi^2-\lambda p_m \phi^3)\phi=0.
\label{41}
\end{equation}
Equations (\ref{38}) and (\ref{41}) are consistent provided $p_m=0$.  
Unlike the Deser's model, here the conformal symmetry is broken down not by adding the {\it ad hoc} mass term $\frac{1}{2}\int\!d^4 x \sqrt{-g} \mu^2 \phi^2$. Rather, it is automatically broken down once an energy-momentum tensor of non-vanishing trace $T^m\neq 0$ is introduced into the action. Therefore, in the specific cosmological conformal frame defined by (\ref{15}), the Einstein equation in the broken phase of conformal symmetry is obtained
\begin{equation}\label{42}
G_{\mu \nu}= 8\pi G T_{\mu \nu}+\tilde{\Lambda} ({g}_{\mu \nu}-2u_{\mu}u_{\nu}),
\end{equation}
where 
\begin{equation}\label{42'}
\tilde{\Lambda}=\frac{9 \lambda}{16\pi G}\rho_m.
\end{equation}
Eq.(\ref{42'}) indicates that $\tilde{\Lambda}$ is a dynamical cosmological term tracking linearly the evolution of matter energy density $\rho_m$.

\subsection{Dynamics of universe in radiation and matter dominant eras}

We take FRW metric (\ref{metric}) and the perfect fluid ${T}^{~r}_{\mu \nu}$
to be substituted in the Einstein equation (\ref{42}). Because of traceless
property $T^{r}=0$, the dynamics of universe in radiation dominant era is exactly the same as discussed in the case of scalar field non-minimally coupled to gravity, in the subsection 3.1.
However, in the matter dominant era we face with new and interesting results.
In fact, once the universe changes its thermodynamical phase from radiation to matter dominance, the conformally invariant action $S[\phi] + S_{r}$ is changed to the conformally non-invariant action $\tilde{S}[\phi] + S_{m}$ due to $T^{m}\neq0$, namely the conformal symmetry is broken down in this phase transition. This phase transition is consistent provided that the pressure is zero. This is an interesting and surprising result because the condition $p_m=0$ is in exact agreement with the current observations on the present status of universe. In other words, only a pressureless matter
can couple consistently to gravity in the conformally non-invariant
action $\tilde{S}[\phi] + S_{m}$.

The filed equations (\ref{42}) subject to $p_m=0$ are then obtained
\begin{equation}
3\frac{\dot{a}^2}{a^2}=8 \pi G \rho_m+\tilde{\Lambda}, \label{44}
\end{equation}
\begin{equation}
\frac{2a\ddot{a}+\dot{a}^2}{a^2}=-\tilde{\Lambda}. \label{45}
\end{equation}
Combining Eqs.(\ref{44}) and (\ref{45}) results in the following acceleration
equation
\begin{equation}
\frac{\ddot{a}}{a}=-\frac{2}{3}(2\pi G \rho_m+\tilde{\Lambda}). \label{46}
\end{equation}
The acceleration is positive provided that
\begin{equation}
8\pi G \rho_m-\frac{9 \lambda \rho_m}{16\pi G}<0. \label{47}
\end{equation}
Bearing in mind that $\rho_m>0$, we find that in order for an accelerating
universe happens in the matter dominant era, the value of self interaction coupling constant $\lambda$ should have the following lower bound   
\begin{equation}
\lambda>\frac{32\pi^2 G^2}{9}. \label{48}
\end{equation}

\subsection{Coincidence problem}

In the subsection 4.3, we studied the coincidence problem and found that
the energy density of matter is of the same order of magnitude as that of the constant vacuum energy density. Although this helps us to theoretically
justify why the order of magnitudes of matter and dark energy densities are the same at present status of the universe, however, the full solution of coincidence problem is achieved if we show that the vacuum energy density, namely the dark energy density, is dynamical and its value is always of the same order of magnitude as that of matter
energy density. To this end, we simply note that the energy density of vacuum or dark energy is given by
\begin{equation}
\rho_V=\frac{\tilde{\Lambda}}{8 \pi G},\label{49}
\end{equation}
which, after using Eqs.(\ref{42'}) and (\ref{48}), yields
\begin{equation}
\rho_V>\frac{1}{4}\rho_m.\label{50}
\end{equation}
This is in good agreement with the current observations on the dark energy
density and shows that the dark energy density is dynamical and follows the evolution of {\it matter} energy density. This solves the coincidence problem. 

\section{Conclusions}

In this paper, we have examined the possibility of having an accelerating
universe in the pressureless matter dominant era by studying a scenario in
which the breakdown of conformal symmetry is responsible for such acceleration.
In this regard, we have considered a self interacting scalar field $\phi$ non-minimally coupled to gravity and matter. The coupling to gravity and matter is established by the terms $R \phi^2$ and $\lambda T \phi^4$, respectively, where $R$ is the Ricci scalar and $T$ is the  trace of energy-momentum tensor. For a traceless energy-momentum tensor like the perfect fluid in radiation
dominant era, the  coupling $\lambda T \phi^4$ vanishes and we are left with
a conformal invariant gravitational action of Deser. It is shown that the universe in this era may experience a decelerating phase in agreement with
observations. After decoupling of matter from radiation and rise of matter dominance, the  coupling $\lambda T^m \phi^4$ no longer vanishes. Then, we face with two surprising results: 1) the conformal symmetry is automatically broken down, 2) the matter automatically becomes pressureless. The corresponding field equations, in the broken phase of conformal symmetry, are obtained and it is shown that the universe may have accelerating phase in matter dominant era, provided the value of self interaction coupling constant $\lambda$ satisfies the lower bound given by Eq.(\ref{48}). Moreover, a reasonable solution to the coincidence problem is introduced.

It is worth noting that the Einstein equation Eq.(\ref{42}) may be rewritten as \begin{equation}
G_{\mu \nu}= 8\pi G T_{\mu \nu}+\tilde{\Lambda} \tilde{g}_{\mu \nu},
\end{equation}
where $\tilde{\Lambda}=-{9 \lambda \rho_m}/{16\pi G}$ and 
\begin{equation}
\tilde{g}_{\mu \nu}=2u_{\mu}u_{\nu}-{g}_{\mu \nu}, 
\end{equation}
which is a metric of Euclidean signature. One may interpret this surprising situation as follows: ``A negative cosmological term can accelerate the universe
in matter dominant era if it is proportional to the matter energy density as $\tilde{\Lambda}=-\alpha \rho_m$ (with $\alpha>2 \pi G$) and is coupled to an Euclidean metric $ \tilde{g}_{\mu \nu}$".

\section*{Acknowledgment}
This work has been supported by a grant/research fund number 
$217/D/5947$ from Azarbaijan Shahid Madani University.

\end{document}